\documentclass[reprint,aps,prd,amsmath,amssymb,showpacs,nofootinbib]{revtex4-1}
\usepackage{hyperref}
\usepackage{epsfig}
\usepackage{graphicx}

\begin{document}

\title{Reconciling the local void with the CMB}

\author{Seshadri Nadathur}
\author{Subir Sarkar}
\affiliation{Rudolf Peierls Centre for Theoretical Physics, University
  of Oxford, Oxford OX1 3NP, UK} \date{\today}

\begin{abstract}
  In the standard cosmological model, the dimming of distant Type Ia
  supernovae is explained by invoking the existence of repulsive `dark
  energy' which is causing the Hubble expansion to accelerate. However
  this may be an artifact of interpreting the data in an
  (oversimplified) homogeneous model universe. In the simplest
  inhomogeneous model which fits the SNe Ia Hubble diagram
  \emph{without} dark energy, we are located close to the centre of a
  void modelled by a Lema\'itre-Tolman-Bondi metric. It has been
  claimed that such models cannot fit the CMB and other cosmological
  data. This is, however, based on the \emph{assumption} of a scale-free
  spectrum for the primordial density perturbation. An alternative
  physically motivated form for the spectrum enables a good fit to
  both SNe Ia (Constitution/Union2) and CMB (WMAP 7-yr) data, and to
  the locally measured Hubble parameter. Constraints from baryon
  acoustic oscillations and primordial nucleosynthesis are also
  satisfied.

\end{abstract}

\maketitle


\section{Introduction}
\label{section:intro}

The simplest cosmological model consistent with the spatial flatness
expectation of inflation is the Einstein-de Sitter (EdS) universe
with $\Omega_m=1$ and $\Omega_K=0$. This formed the basis of
the ``standard cold dark matter'' (SCDM) cosmology which provided a good
description of the large-scale structure of the universe
\cite{Blumenthal:1984bp,Davis:1985rj}. It was noted however that it is
inconsistent with the angular power spectrum of the clustering of
galaxies in the APM survey \cite{Efstathiou:1990xe}. Subsequently
large-angle anisotropies in the cosmic microwave background (CMB) were
detected by COBE \cite{Smoot:1992td}, thus providing an absolute
normalization of the amplitude of primordial density
perturbations. The SCDM model assumes that the power spectrum of the
primordial density perturbations has the scale-invariant form: $P(k)
\equiv |\delta_k|^2 \propto k^n$, with $n=1$ and the predicted
amplitude of matter fluctuations on small (cluster and galaxy) scales
is then too high relative to observations \cite{White:1992ri}. However
if the epoch of matter-radiation equality is delayed by lowering the
CDM density to $\Omega_m \sim 0.3$, then the peak in the
power spectrum of density fluctuations is shifted to larger scales
thus decreasing the power on small scales and enabling a match to the
data. To maintain spatial flatness a non-zero value of the
cosmological constant was invoked, with $\Lambda \sim 2H_0^2$
corresponding to $\Omega_\Lambda \equiv \Lambda/3H_0^2 \sim
0.7$. Subsequently it was also observed that Type Ia supernovae
(SNe~Ia) at redshift $z \simeq 0.5$ appear $\sim25\%$ fainter than
expected in an EdS universe
\cite{Riess:1998cb,Perlmutter:1998np}. Together with measurements of
galaxy clustering in the 2dF survey \cite{Efstathiou:2001cw} and of
cosmic microwave background (CMB) anisotropies by WMAP
\cite{Spergel:2003cb}, this changed the standard cosmological model
to an accelerating universe with a dominant cosmological constant
term, which has been widely interpreted as a manifestation of the
physical vacuum or ``dark energy''. This ``concordance'' $\Lambda$CDM
cosmology (with $\Omega_\Lambda \simeq 0.7$, $\Omega_m \simeq
0.3$, $h \simeq 0.7$) has proved to be consistent with other
cosmological data, in particular, baryonic acoustic oscillations
detected in the SDSS \cite{Eisenstein:2005su} and measurements of mass
fluctuations from clusters and weak lensing
\cite{Contaldi:2003hi}. Further observations of both SNe~Ia
\cite{Riess:2004nr,Astier:2005qq,WoodVasey:2007jb} and the WMAP 3-year
results \cite{Spergel:2006hy} have continued to firm up the model.

Embarrassingly, however, this model lacks a {\em physical} basis. There
are two serious problems with the notion that the universe is dominated by
some form of vacuum energy. The first is the notorious fine-tuning
problem of vacuum fluctuations in quantum field theory --- the energy
scale of the inferred cosmological energy density is $\sim10^{-12}$
GeV, which is many orders of magnitude below the energy scale of $\sim10^2$ GeV
of the Standard Model of particle physics, not to mention the Planck
scale of $M_\mathrm{P}\equiv (8\pi
G_\mathrm{N})^{-1/2}\simeq2.4\times10^{18}$ GeV
\cite{Weinberg:1988cp}. The second is the equally acute coincidence
problem: since $\Omega_\Lambda/\Omega_m$ evolves as the cube
of the cosmic scale factor $a(t)$, there is no reason to expect it to be
of ${\cal O}(1)$ {\em today}, yet this is supposedly the case.

Interestingly the WMAP results alone do not require dark energy if the
assumption of a scale-invariant primordial power spectrum is relaxed.
This is well justified given our present ignorance of the physics
underlying inflation which is believed to have created these
fluctuations. It has been demonstrated \cite{Hunt:2007dn,Hunt:2008wp}
that the temperature angular power spectrum of an EdS universe with
$h \simeq 0.44$ matches the WMAP data well if the primordial power is
enhanced by $\sim 30\%$ in the region of the second and third acoustic
peaks (corresponding to spatial scales of $k \sim
0.01-0.1h~\mathrm{Mpc}^{-1}$). This alternative model with {\em no}
dark energy has a slightly better $\chi^2$ for the fit to WMAP-3 data
than the concordance ``power-law $\Lambda$CDM model'' and, inspite of
having more parameters, has an {\em equal} value of the Akaike
information criterion used in model selection. Other EdS models with
a broken power-law spectrum \cite{Blanchard:2003du} have also been
shown to fit the WMAP data.  Moreover, an EdS universe can fit
measurements of the galaxy power spectrum if it includes a $\sim 10\%$
component of hot dark matter in the form of massive neutrinos of mass
$\sim 0.5$~eV \cite{Blanchard:2003du,Hunt:2007dn,Hunt:2008wp}. Clearly
the main evidence for dark energy comes from the SNe~Ia Hubble
diagram.

It should be kept in mind that the acceleration of the expansion rate
is not directly measured but inferred from measurements of the
apparent magnitudes and redshifts of SNe~Ia. Indeed \emph{all} the
evidence for dark energy is geometrical, i.e., based on
interpreting the data in an assumed homogeneous model universe.
So far there is no convincing observation of dynamical manifestations,
e.g., the ``late integrated Sachs-Wolfe effect''. In fact what
is actually inferred from observations is {\em not} an energy density,
just a value of ${\cal O}(H_0^2)$ for the otherwise unconstrained
$\Lambda$ term in the Friedmann equation. It has been suggested that
this may simply be an artifact of interpreting imprecisely measured
cosmological data in the oversimplified framework of an universe
assumed to be described by the exactly isotropic and homogeneous
Friedmann-Robertson-Walker (FRW) metric, in which $H_0 \sim 10^{-42}
\mathrm{GeV} \sim (10^{28} \mathrm{cm})^{-1}$ is the {\em only} scale
 \cite{Sarkar:2007cx}. Note that the non-zero value of
$\Omega_\Lambda \equiv \Lambda/3H_0^2$ is inferred from the ``cosmic
sum rule'' $\Omega_m + \Omega_K + \Omega_\Lambda=1$, which is
just a restatement of the Friedmann equation. If however this equation
does not describe the real universe exactly, and in order to do so
\emph{other} non-zero terms ought to have been added to the sum rule, then we
may mistakenly infer a value for $\Omega_\Lambda$ of ${\cal O}(1)$ if
these other terms are in fact important. For example, in an
inhomogeneous universe averaged quantities satisfy modified Friedmann
equations which contain extra terms since the operations of spatial
averaging and time evolution do not commute
\cite{Buchert:1999er}. These ``backreaction'' terms depend upon the
variance of the local expansion rate and hence increase as
inhomogenities develop. However although backreaction behaves just
like a cosmological constant, whether its expected magnitude can
indeed account for the apparent cosmological acceleration is debated
and remains an open question at present
\cite{Wetterich:2001kr,Ishibashi:2005sj,Vanderveld:2007cq,Wiltshire:2007fg,Khosravi:2007bq,Leith:2007ay,Behrend:2007mf,Rasanen:2008it,Li:2008yj,Paranjape:2008jc}.

Another possibility is that inhomogeneities affect light propagation
on large scales and cause the luminosity distance-redshift relation to
resemble that expected for an accelerating universe. This has been
investigated for a ``Swiss-cheese'' universe in which voids modelled by
patches of Lema\'{i}tre-Tolman-Bondi (LTB) space-time are distributed
throughout a homogenous background. However, the results depend on the
specific model: some authors find the change in light propagation to
be negligible because of cancellation effects
\cite{Biswas:2007gi,Brouzakis:2007zi,Brouzakis:2008uw,Valkenburg:2009iw},
whereas others claim it can partly mimic dark energy
\cite{Marra:2007pm,Marra:2007gc,Kainulainen:2009sx}. It may be that
observers preferentially choose sky regions with underdense
foregrounds when studing distant objects such as SNe~Ia, so the
expansion rate along the line of sight is then greater than average;
such a selection effect may also allow an inhomogeneous universe to
fit the observations without dark energy \cite{Mattsson:2007tj}.
   
In this paper we are mainly interested in a ``local void'' (sometimes
referred to as ``Hubble bubble'') as an explanation for dark energy;
to prevent an excessive CMB dipole moment due to our peculiar velocity
we must be located near the centre of the void. An underdense void
expands faster than its surroundings, thus younger supernovae inside
the void would be observed to be receding more rapidly than older
supernovae outside the void. Under the assumption of homogeneity this
would lead to the mistaken conclusion that the expansion rate of the
universe is accelerating, although both the void and the global
universe are actually decelerating.  The local void scenario has been
investigated by several authors using a variety of methods
\cite{Moffat:1994qy,Tomita:1999rw,Celerier:1999hp,Tomita:1999qn,Tomita:2000rf,Tomita:2000jj,Tomita:2001gh,Iguchi:2001sq,Tomita:2002df,Moffat:2005yx,Moffat:2005ii,Alnes:2005rw,Mansouri:2005rf,Vanderveld:2006rb,Garfinkle:2006sb,Chung:2006xh,Biswas:2006ub,Alnes:2006uk,Caldwell:2007yu,Alexander:2007xx,Clarkson:2007pz,Uzan:2008qp,GarciaBellido:2008nz,GarciaBellido:2008gd,Clifton:2008hv,Bolejko:2008cm}.
By modelling the void as a open FRW
region joined by a singular mass shell to a FRW background, it was
found that a void with radius 200 Mpc can fit the supernova Hubble
diagram without dark energy \cite{Tomita:2001gh}. It was also shown
that a LTB region which reduces to a EdS cosmology with $h=0.51$ at a
radius of 1.4~Gpc can match both the supernova data and the location
of the first acoustic peak in the CMB
\cite{Alnes:2005rw}. Ref.\cite{Alexander:2007xx} attempted to find the
smallest possible void consistent with the current supernova results
--- their LTB-based `minimal void' model has a radius of 350
Mpc. Unfortunately, since this model is equivalent to an EdS universe
with $h=0.44$ {\em outside} the void where the SDSS luminous red
galaxies lie, as it stands it is unable to fit the measurements of the
baryonic acoustic oscillation (BAO) peak at $z\sim0.35$
\cite{Blanchard:2005ev}. LTB models of much larger voids were
considered in Ref.\cite{GarciaBellido:2008nz} (with radii of 2.3 Gpc
and 2.5 Gpc and Hubble contrasts of 0.18 and 0.30 respectively) and it
was demonstrated they can fit the BAO data, as well as the SNe\,Ia
data and the location of the first CMB peak. Ref.\cite{Clifton:2008hv}
found the best fit to the SNe Ia data for a void of radius $1.3 \pm
0.2$ Gpc and Ref.\cite{Bolejko:2008cm} confirmed that such a void
provides an excellent fit to the ``Union'' dataset of SNe\,Ia.

In this paper we demonstrate that, contrary to the results obtained in Refs.~\cite{Clifton:2009kx,Zibin:2008vk,Moss:2010jx,Biswas:2010xm}, a Gpc-sized void can simultaneously fit the SNe Ia data as well as the full CMB power spectrum, while also satisfying constraints from local Hubble measurements, primordial nucleosynthesis and the BAO data, if the primordial power spectrum is \emph{not} assumed to be nearly scale-invariant. The layout of the paper is as follows. In Sec.~\ref{section:LTB} we summarize the general relativistic framework for LTB models and describe the characterization of the void. In Sec.~\ref{section:primordial} we discuss the form of the primordial power, and present a physical model with a primordial power spectrum that is not scale-free. In order to compare observables in the void model to existing cosmological data, some formalism needs to be developed. This is done in Sec.~\ref{section:fitting}, and the statistical approach is discussed in Sec.~\ref{section:method}. Finally Sec.~\ref{section:results} presents the main results of the paper.

\section{LTB void models}
\label{section:LTB}

\subsection{The metric and solution}
\label{subsection:metric}
We model the void as an isotropic, radially inhomogeneous universe
described by the LTB metric:
\begin{equation}
ds^2 = -c^2dt^2 + \frac{A^{\prime2}(r, t)}{1 + K(r)} dr^2 
 + A^2(r, t)d\Omega^2,
\end{equation}
where a prime denotes the partial derivative with respect to
coordinate distance $r$, and the curvature $K(r)$ is a free function,
bounded by $K<1$. This reduces to the usual FRW metric in the limit
where $A(r,t)\rightarrow a(t)r$ and $K(r)\rightarrow\kappa r^2$.

We define two Hubble rates:
\begin{equation}
H_{\perp} \equiv \frac{\dot{A}(r,t)}{A(r,t)}, \:\: 
H_{\parallel}\equiv\frac{\dot{A}^\prime(r, t)}{A^\prime(r, t)},
\end{equation} 
where an overdot denotes the partial derivative with respect to
$t$. The analogue of the Friedmann equation is
\begin{equation}
H_\perp^2 = \frac{F(r)}{A^3(r, t)} + \frac{c^2K(r)}{A^2(r, t)},
\end{equation}
where $F(r)>0$ is another free function which determines the local
energy density through
\begin{equation}
8\pi G\rho(r, t) = \frac{F^\prime(r)}{A^2(r, t)A^\prime(r, t)}. 
\end{equation}
We define dimensionless density parameters $\Omega_M(r)$ and
$\Omega_K(r)$ such that
\begin{equation}
F(r) = H_0^2(r)\Omega_M(r)A_0^3(r),
\end{equation}
and
\begin{equation}
c^2K(r) = H_0^2(r)\Omega_K(r)A_0^2(r),
\end{equation}
where $H_0(r)$ and $A_0(r)$ are the values of $H_\perp(r,t)$ and
$A(r,t)$ respectively at the present time $t=t_0$. The Friedmann
equation then becomes \cite{Enqvist:2006cg}:
\begin{equation}
\label{Friedmann}
H_\perp^2=H_0^2\left[\Omega_M\left(\frac{A_0}{A}\right)^3 + \Omega_K\left(\frac{A_0}{A}\right)^2\right],
\end{equation}
so $\Omega_M(r)+\Omega_K(r)=1$. This equation can be
integrated from the time of the Big Bang,
$t_\mathrm{B}=t_\mathrm{B}(r)$, to yield the age of the universe at
any given $(r, t)$:
\begin{equation}
t - t_\mathrm{B}(r) = \frac{1}{H_0(r)} \int_0^{A/A_0} 
 \frac{\mathrm{d}x}{\sqrt{\Omega_M(r) x^{-1} + \Omega_K(r)}}.
\end{equation}
We thus have two functional degrees of freedom, in
$\Omega_M(r)$ and $t_\mathrm{B}(r)$, which can be chosen as
desired. (The third function, $A_0(r)$, corresponds to a gauge mode
and we choose to set $A_0(r)=r$.) A spatially varying
$t_\mathrm{B}$ corresponds to a decaying mode \cite{Zibin:2008vj},
so for simplicity we set $t_\mathrm{B}=0$ everywhere, so that at
the current time $t_0$:
\begin{eqnarray}
\label{H0}
H_0(r) = \left\{
\begin{array}{cr}
  \displaystyle \frac{- \sqrt{-\Omega_K }+\Omega_M\sin^{-1}\sqrt{-\frac{\Omega_K}{\Omega_M}} } {t_0\left(-\Omega_K\right)^{3/2}}\,,   &  \Omega_K<0; \\[4mm]
  \displaystyle  \frac{2}{3t_0},   & \Omega_K=0; \\[2mm]
  \displaystyle  \frac{\sqrt{\Omega_K } - \Omega_M\sinh^{-1}\sqrt{\frac{\Omega_K}{\Omega_M}} } {t_0\Omega_K^{3/2}},   &  \Omega_K>0.
\end{array}
\right.
\end{eqnarray}
The void model can then be specified by the choice of one free
function, which we take to be $\Omega_M(r)$, and a constant
$H\equiv H_0(0)$ which determines the \emph{local} Hubble rate at the
centre (and is equivalent to choosing $t_0$). Note that both at the
centre of the void and far outside the void, the definition of
$\Omega_M(r)$ reduces to the standard FRW density parameter
$\Omega_m$.

The solution to Eq.(\ref{Friedmann}) for general $r$ and $t$ can be
given in parametric form for the different values of $\Omega_K(r)$ (or
$K(r)$) as follows \cite{Celerier:1999hp}:
\begin{itemize}
\item for $\Omega_K(r)>0$:
\begin{subequations}
\label{Omk>0}
\begin{eqnarray}
A&=&\frac{\Omega_M(r)A_0(r)}{2\Omega_K(r)}\left(\cosh\eta-1\right), \label{Omk>0:1} \\
H_0t&=&\frac{\Omega_M(r)}{2\Omega_K^{3/2}(r)}\left(\sinh\eta-\eta\right).\label{Omk>0:2} 
\end{eqnarray}
\end{subequations}
\item for $\Omega_K(r)=0$:
\begin{equation}
\label{Omk=0}
A=\frac{1}{2}\left(18\Omega_M(r)\right)^{1/3}\left(H_0t\right)^{2/3}A_0(r).
\end{equation}
\item for $\Omega_K(r)<0$:
\begin{subequations}
\label{Omk<0}
\begin{eqnarray}
A&=&\frac{\Omega_M(r)A_0(r)}{2\vert\Omega_K(r)\vert}\left(1-\cos u\right),\label{Omk<0:1} \\
H_0t&=&\frac{\Omega_M(r)}{2\vert\Omega_K(r)\vert^{3/2}}\left(u-\sin u\right). \label{Omk<0:2} 
\end{eqnarray}
\end{subequations}
\end{itemize} 

Light travels to an observer at the centre of the void along null
radial incoming geodesics described by \cite{Celerier:1999hp}:
\begin{subequations}
\begin{eqnarray}
\label{geodesics}
\frac{dt}{dz}&=&-\frac{1}{\left(1+z\right)H_\parallel(z)} ,\label{geodesics:t}\\
\frac{dr}{dz}&=&\frac{c\sqrt{1+K\left(r\right)}}{\left(1+z\right)A^\prime(z)H_\parallel(z)}, 
\label{geodesics:r}
\end{eqnarray}
\end{subequations}
where $H_\parallel(z)=H_\parallel\left(r(z),t(z)\right)$
\emph{etc}. The angular diameter distance at redshift $z$ is then
given by
\begin{equation}
\label{d_A}
d_A(z)=A\left(r(z),t(z)\right) ,
\end{equation}
and the luminosity distance by
\begin{equation}
\label{d_L}
d_\mathrm{L}(z)=\left(1+z\right)^2A\left(r(z),t(z)\right) .
\end{equation}
All observable quantities along the light cone can be calculated from these equations.

\subsection{Void profile}
\label{subsection:profile}

We can now choose the void profile by specifying
$\Omega_M(r)$ and the local value of the Hubble rate by
specifying $H$. Although the void profile may have any shape, we
restrict ourselves to the simple Gaussian form:
\begin{equation}
\label{Om(r)}
\Omega_M(r)=\Omega_\mathrm{out}-\left(\Omega_\mathrm{out}-\Omega_\mathrm{in}\right)\exp\left[-\left(\frac{r}{r_0}\right)^2\right],
\end{equation}
where $\Omega_\mathrm{in}$ and $\Omega_\mathrm{out}$ correspond to the
matter density parameter at the centre of the void and at infinity,
respectively, and $r_0$ characterizes the width of the void. We wish
to look only at voids that are asymptotically EdS, so we set $\Omega_\mathrm{out}=1$. Thus distances in the void model are completely specified by the three parameters
$\Omega_\mathrm{in}$, $r_0$ and $H$.

We wish to stress that in restricting ourselves to voids which have a
Gaussian profile, we may be missing the model that fits the data
best. In principle we could sample a wider class of profiles and
choose the form that gives the best fit. However in this paper we are
concerned mainly with providing a counter-example of a void which can
simultaneously fit both the SNe Ia magnitudes and the CMB spectrum, so
we do not perform this search.

\section{Primordial power spectra}
\label{section:primordial}

The observed power spectrum of CMB anisotropies in any cosmological
model is a convolution of three unknowns. The first two are the local
physics at the time of recombination, which is dependent on the
composition of the universe at that time, and the angular diameter
distance to the last scattering surface (LSS), which depends on the
geometry of the universe. Both of these are completely specified by
the choice of the void model as described in
Section~\ref{subsection:profile}, along with the further specification
of the baryon fraction, $\Omega_\mathrm{b}$ and the baryon-to-photon
ratio $\eta \equiv n_\mathrm{b}/n_\gamma$.

The third unknown is the shape of the primordial power spectrum of
density perturbations. In toy models of inflation this is close to
scale-invariant and featureless. Under the assumption that the
primordial spectrum is described by a simple power law, the standard
$\Lambda$CDM concordance cosmological model fits the observed angular
power spectrum reasonably well \cite{Komatsu:2010fb}. However, there
is no independent evidence for this form of the primordial power and
as the observed anisotropies arise as a convolution of the
\emph{assumed} primordial spectrum with the transfer function of the
\emph{assumed} cosmological model, it is clear that we cannot
determine one without making assumptions about the other. (In fact,
there are indications that the primordial spectrum is \emph{not}
scale-free, even when a $\Lambda$CDM cosmology is assumed
\cite{Martin:2003sg,Kogo:2004vt,Kogo:2005qi,Shafieloo:2006hs,Shafieloo:2003gf,
  TocchiniValentini:2005ja,TocchiniValentini:2004ht,Nicholson:2009zj,Nicholson:2009pi,Ichiki:2009zz}.)

In Refs.~\cite{Clifton:2009kx,Zibin:2008vk,Moss:2010jx,Biswas:2010xm}
it is argued that, \emph{assuming near-scale invariance of the
  primordial power}, void models cannot simultaneously provide an
explanation for supernovae magnitudes and fit the observed CMB
spectrum, unless the void is extremely deep \cite{Clifton:2009kx} or
embedded in a universe with large non-zero overall curvature
\cite{Biswas:2010xm}. However, in considering void models as an alternative
to dark energy, we are in any case
departing from the concordance cosmology, so there is no need to
retain the assumption that the primordial power spectrum is
scale-free. If this assumption is relaxed then even an EdS cosmology {\em without} dark energy can fit the CMB data \cite{Blanchard:2003du,Hunt:2007dn}. In this paper we
consider a particular alternative form of
the primordial $\mathcal{P}(k)$ that is not scale-free but is in fact
physically well-motivated, as described below.

\subsection{Bump model}
\label{subsection:bump}

Whereas the simplest toy models of inflation contain only a single
scalar field which rolls slowly down its potential, \emph{physical}
models generically contain other fields, whose evolution is typically
not slow-roll. These fields may couple to the inflaton and
affect its evolution, thus breaking the scale-free nature of the
primordial power, with important consequences.

An example of such a physical model is ``multiple inflation''
\cite{Adams:1997de} in the framework of $N=1$ supergravity, the
locally realised version of supersymmetry (SUSY). This model includes
``flat-direction'' fields $\psi$ which have gauge and Yukawa couplings
to ordinary matter but are only gravitationally coupled to the
inflaton. Such flat-directions have been classified and tabulated in
the Minimal Supersymmetric Standard Model (MSSM)
\cite{Gherghetta:1995dv}. As the universe cools during inflation, the
$\psi$ fields undergo symmetry-breaking phase transitions, causing a
sudden change in the effective mass of the inflaton (which is assumed
to be a field in a ``hidden sector''). For a single flat-direction
field, this produces a `step' in the spectrum of the curvature
perturbation \cite{Hunt:2004vt}. If more than one flat-direction field
is present, they can couple to the inflaton with opposite signs, and
instead produce a `bump' feature in the power spectrum
\cite{Hunt:2007dn}. (A toy model that produces a similar ``step''
feature in the power spectrum has also been proposed
\cite{Adams:2001vc} and has been studied with respect to fitting the
WMAP 1-year \cite{Peiris:2003ff,Covi:2006ci} and 3-year \cite{Hamann:2007pa}
data, albeit in a different context to that considered here. Other
signatures of multiple inflation, in particular the generation of
associated non-Gaussianities, have also been studied
\cite{Hotchkiss:2009pj}.)

The potential for the inflaton $\phi$ and the flat-direction fields
$\psi_i$ is then given by
\begin{eqnarray}
V(\phi,\psi_1,\psi_2)=&&V_0-\frac{1}{2}m^2H_\mathrm{I}^2\phi^2+\frac{1}{2}\lambda_1H_\mathrm{I}^2\phi^2\psi_1^2 \nonumber\\ 
&&-\frac{1}{2}\mu_1^2H_\mathrm{I}^2\psi_1^2+\gamma_1\psi_1^{n_1}+\frac{1}{2}\lambda_2H_\mathrm{I}^2\phi^2\psi_2^2 \nonumber\\ 
\label{multinfpotential}
&&-\frac{1}{2}\mu_2^2H_\mathrm{I}^2\psi_2^2+\gamma_2\psi_2^n\,,
\end{eqnarray}
where $mH_\mathrm{I}$ and $\mu_iH_\mathrm{I}$ are the masses of the
$\phi$ and $\psi_i$ fields respectively, $\lambda_iH_\mathrm{I}^2$ is
the coupling of the $\psi_i$ field to the inflaton, $\gamma_i$ is the
coefficient of the non-renormalizable operator of order $n_i$ which
lifts the potential of the $\psi_i$ field and $H_\mathrm{I}$ is the
Hubble scale during inflation (units of Planck mass $M_\mathrm{P}=1$
are used throughout). The two flat-direction fields remain trapped at
the origin by thermal effects until the phase transitions take place
at times $t_i$. Natural values for the parameters of this model are
$\vert\lambda_i\vert\sim1$, $\mu_i^2\sim3$, and
$\gamma_i\sim1$. According to the list of flat-direction fields in
Ref.~\cite{Gherghetta:1995dv}, the ones with $n_i=12,16$ will be the
most relevant. It is usually assumed that some symmetry protects the
inflaton mass $m$ from SUSY-breaking corrections (the ``$\eta$
problem''), thus allowing sufficient e-folds of inflation to
occur. Note that the times of the phase transitions $t_1$ and $t_2$
--- or, equivalently, the scales $k_1$ and $k_2$ at which the effects
of the flat-direction fields begin to be felt --- are however
arbitrary.

In Ref.~\cite{Hunt:2007dn} a full likelihood analysis was performed to
demonstrate that a primordial power spectrum with such a `bump' would
allow even an EdS universe with $\Lambda = 0$ to fit the WMAP 3-year
data, although this requires a rather low value of $h_0 \simeq
0.44$. As void models generically imply a lower global value of $h_0$
than the local value at the centre where we are located, such a
primordial power spectrum could help a void model to fit the CMB.

The primordial power spectrum produced by this `bump' model can be
calculated exactly by numerically solving the equation of motion of
the $\phi$ field according to the potential in
Eq.(\ref{multinfpotential}). However, we do not choose to do so for
two reasons. Firstly, this evaluation is computationally expensive and
slows an MCMC analysis considerably. Secondly, our results
regarding void models are dependent only on the existence of such a
feature, rather than the details of the particular theory that
explains it. Multiple inflation provides an example of such a theory,
but similar features with steps and oscillations may also be produced,
\emph{e.g.}, in DBI inflation \cite{Brax:2010tq}. Strictly speaking it is not
even necessary to assume that these features arise from an
inflationary model at all.

\begin{figure}[htbp]
\includegraphics[scale=0.72]{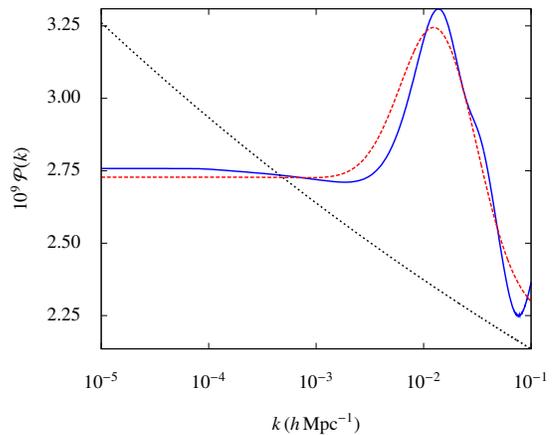}
\caption{\label{fig:Pk} The solid (blue) curve shows the primordial
  power spectrum obtained from a full calculation of the multiple
  inflation model, with parameters as described in the text. The
  dashed (red) curve shows the simple parameterization (\ref{Pkfeat}),
  with parameter values $\mathcal{P}_0=2.48\times10^{-9}$, $a=0.1$,
  $b=2.0$, $c=3.0$ and $k_0=0.015$. For comparison, the dotted (black)
  curve shows a standard power-law spectrum with slope
  $n_\mathrm{s}=0.954$, amplitude $A_\mathrm{S}=2.3\times10^{-9}$,
  which are the best-fit values for the $\Lambda$CDM model with no
  ``running'' to the CMB+$\mathrm{SN}_\mathrm{CfA}$ datasets
  (Sec.~\ref{section:results}).}
\end{figure}

With this in mind, we introduce a parameterization of the primordial
scalar power $\mathcal{P}(k)$ to capture the essential features of the
`bump' model instead of attempting to reproduce it exactly:
\begin{equation}
\label{Pkfeat}
\mathcal{P}(k)=\mathcal{P}_0\left(1-a\tanh(bx)+c\mathrm{e}^{-(bx)^2}\right)\,,
\end{equation}
where $x\equiv\log_{10}(k/k_0)$. In Fig.~\ref{fig:Pk} we plot
$\mathcal{P}(k)$ obtained from a full calculation in the ``bump'' model,
with parameter values $n_1=16$, $n_2=12$, $\mu_1^2=\mu_2^2=3$,
$\gamma_1=\gamma_2=1$, $\lambda_1=-0.2$, $\lambda_2=1$ and $m^2=0.05$
(all set at their ``natural'' values, save for $\lambda_1$). $H_\mathrm{I}$,
$k_1$ and $k_2$ are free parameters and are chosen so as to broadly
reproduce the best-fit primordial spectrum from
Ref.~\cite{Hunt:2007dn}. In the same figure we also plot the form of
our simple parameterization. We do not consider any tensor modes, as
in ``small field'' models such as multiple inflation these are always
negligibly small.

This form of $\mathcal{P}(k)$ is thus determined by 5 free parameters:
$\mathcal{P}_0$, $a$, $b$, $c$ and $k_0$. This may appear to be a step
backwards from the standard power-law form, which depends on only 3
($A_\mathrm{S}$, $n_\mathrm{s}$ and $n_\mathrm{run}$ --- the
amplitude, slope and ``running'' of the spectrum). However it is
important to bear in mind that we are merely using an empirical
parameterization for a $\mathcal{P}(k)$ that is ultimately the result
of some underlying theory, in which all of these parameters are in
fact determined by fundamental physics.

On the other hand, it can be argued that the standard parameterization
of a power-law hides other parameters which are artificially set to
zero through the overriding assumption that only a single scalar field
is involved. In the absence of an accepted physical mechanism for
inflation, we feel that all plausible alternatives ought to be
considered, rather than judging purely on the basis of naive
parameter-counting. It is with this rationale that we use this form of
the primordial power in constructing our counter-example.


\section{Fitting the model to observations}
\label{section:fitting}
In this section we will discuss how to fit the void model to the best
avaliable cosmological data: the WMAP 7-year release
\cite{Komatsu:2010fb}, SNe Ia data from the Constitution
\cite{Hicken:2009dk} and the Union2 compilation
\cite{Amanullah:2010vv}, the local Hubble rate measurement by the
Hubble Key Project \cite{Freedman:2000cf}, constraints from big bang
nucleosynthesis \cite{Nakamura:2010zzi}, and BAO measurements from SDSS DR7 \cite{Percival:2009xn}.

As cosmology in an LTB metric differs from the standard FRW approach,
we need to develop some formalism to allow us to confront these
datasets in a consistent manner. This is outlined in the following.

\subsection{SNe Ia magnitudes}
\label{subsection:SN}

The distance modulus of a supernova is defined as the residual between
its apparent and absolute magnitudes $m$ and $M$, and is related to its
luminosity distance as
\begin{equation}
\mu=m-M=5\log_{10}\left[\frac{d_\mathrm{L}}{\mathrm{Mpc}}\right]+25\,,
\end{equation}
so that, knowing $\mu$ for each object in a dataset, we can use
Eq.(\ref{d_L}) to compare a void model to the data.

We use two datasets: the Constitution compilation of 397 SNe in the
redshift range $z=0.015-1.55$ \cite{Hicken:2009dk} and the Union2
compilation of 557 SNe in the range $z=0.015-1.4$
\cite{Amanullah:2010vv}. The Constitution sample uses the SALT
lightcurve fitter while Union2 uses the newer SALT2 fitter. Neither
fitter, however, directly provides the value of $\mu$ for each object
--- for instance, the SALT fitter provides as output values of
$m_B^\mathrm{max}$ (the rest-frame peak magnitude in the B-band), a
time stretch factor, $s$, and a colour parameter, $c$, for each
supernova, from which the distance modulus is calculated as:
\begin{equation}
\mu_i=m^\mathrm{max}_{B,i}-M+\alpha\cdot\left(s_i-1\right)-\beta\cdot c_i,
\end{equation}
where $M$, $\alpha$ and $\beta$ are empirical coefficients whose
values are determined by marginalizing over the fit to a particular
fiducial cosmology. The SALT2 fitter follows the same principle,
though with a `stretch factor', $x_1$, that is analogous to but not
the same as $s$.

Both the Constitution and Union2 compilations provide tabulated values
of distance modulus $\mu$ and its estimated error $\sigma_\mu$ for a
choice of coefficients $M$, $\alpha$ and $\beta$ calculated for the
best-fit flat $\Lambda$CDM model. In addition, $\sigma_\mu$ includes
an important contribution from a systematic uncertainty whose value is
\emph{chosen} such that the reduced $\chi^2$ value for the flat
$\Lambda$CDM model is of order unity. Hence when comparing any
alternative model to $\Lambda$CDM on the basis of their respective
fits to the SNe Ia data, we ought to in principle perform the entire
fitting analysis from scratch, choosing the values of $M$, $\alpha$
and $\beta$ that produce the best fit for the particular model under
consideration, and adjusting the systematic error inserted by hand in
an equivalent manner. However, to do so we would require not only the
actual lightcurve fitter outputs for each supernova (which the
publicly downloadable data for the Constitution set does contain, but
Union2 does not), but also the full covariance matrix of the
lightcurve fits, which neither compilation provides.\footnote{This is
  provided in the SNLS 3-year data release \cite{Guy:2010bc}; however
  it contains only 231 objects, so we do not consider it here.}
Therefore, we are forced to adopt the common (if incorrect) procedure
of fitting the void model by marginalizing over the unknown absolute
magnitude $M$ alone, while bearing in mind the caveat that this may
bias our results and our best-fit void parameters in an unknown
manner.

\subsection{CMB power spectrum}
\label{subsection:CMB}

It has been shown in several previous studies that void models are
only very loosely constrained by the position of the first peak of the
CMB \cite{Alnes:2005rw,GarciaBellido:2008nz}. However, the WMAP
satellite has in fact measured the angular power spectrum
$\mathit{C}_l$s over a wide range of multipoles, $l$, and it is
preferable to use as much of this data as possible. This entails the
calculation of the full predicted $\mathit{C}_l$ spectrum for the void
model. This has been done using various methods
\cite{Clifton:2009kx,Zibin:2008vk,Moss:2010jx,Biswas:2010xm}, but
always with the assumption of a nearly scale-invariant primordial
spectrum.

In order to be able to calculate the predicted angular power spectrum
for an LTB model using one of the publically available Boltzmann codes
(we use a version of CAMB \cite{Lewis:1999bs}, modified to accept the
primordial power spectra we consider), we use a version of the
``effective EdS approach''
\cite{Zibin:2008vk,Moss:2010jx}. In brief, this consists of
constructing an effective EdS model which has the same physics at
recombination as the given void model, and the same angular diameter
distance to the LSS. Then, given the same primordial power spectrum,
the $\mathit{C}_l$s of the effective model and the void model will be
the same, at intermediate and small angular scales. Thus the effective
model can be used as a calculational tool to obtain the power spectrum
for the void model under consideration. Note however that the
effective EdS model will in general have a central temperature,
$T_0^\mathrm{EdS}$, and a central Hubble rate, $H_0^\mathrm{EdS}$,
that are {\em different} from the actual $T_0$ and $H_0$ that we
observe today. This is because the physics has been matched at early
times; at late times the models must then necessarily differ.

To generate the effective model, we adopt the following
procedure. First we specify the void profile and local Hubble rate by
choosing the parameters $\Omega_\mathrm{in}$, $r_0$ and $H$. Then,
using Eqs. \ref{geodesics:t}) and (\ref{geodesics:r}) we numerically
integrate out from the centre of the void along the past light cone to
obtain the coordinates $\left(r_m,t_m\right)$ at an intermediate
redshift $z_m$. As in Refs.~\cite{Zibin:2008vk,Moss:2010jx} we choose
this redshift to be $z_m=100$, where the spatial curvature of the void
is negligible but the radiation density is still small and can
justifiably be ignored in the calculations.

At these coordinates we now calculate the Hubble rate,
$H_m=H_{\perp/\parallel}\left(r_m,t_m\right)$ and the angular diameter
distance $A_m=A\left(r_m,t_m\right)$. Let us denote by
$r^\mathrm{EdS}\left(t\right)$ the comoving radial coordinate of a
radial light ray in the EdS universe and by $z^\mathrm{EdS}$ the
redshifts seen by an observer at $r^\mathrm{EdS}=0$. To ensure the
matching of the distances to the LSS, we impose the condition
$A_m=a(r_m^\mathrm{EdS})r_m^\mathrm{EdS}$ for EdS scale factor
$a$. This provides us with the relation
\begin{equation}
\label{Am}
A_m=\frac{2c}{\left(1+z_m^\mathrm{EdS}\right)H_0^\mathrm{EdS}}\left(1-\frac{1}{\sqrt{1+z_m^\mathrm{EdS}}}\right),
\end{equation}
where $z_m^\mathrm{EdS}$ is the redshift in the EdS universe at
coordinates $\left(r_m^\mathrm{EdS},t_m\right)$. (Note that our
procedure implies $z_m^\mathrm{EdS}\neq z_m$ and thus differs slightly
from the equivalent method outlined in
Refs.\cite{Zibin:2008vk,Moss:2010jx} where instead
$z_0^\mathrm{EdS}\neq0$.) To ensure the same physics at early times in
the two models we also match the Hubble rates,
$H_m=H^\mathrm{EdS}\left(z_m^\mathrm{EdS}\right)$, and this combined
with Eq.(\ref{Am}) provides us with an expression for
$z_m^\mathrm{EdS}$
\begin{equation}
\label{zmEdS}
z_m^\mathrm{EdS}=\frac{A_m^2H_m^2+4cA_mH_m}{4c^2}.
\end{equation}
Given the value of $z_m^\mathrm{EdS}$, we can then calculate the
effective EdS mean temperature and Hubble rate via
\begin{subequations}
\begin{eqnarray}
\label{T0EdS,H0EdS}
T_0^\mathrm{EdS}&=&T_0\left(\frac{1+z_m}{1+z_m^\mathrm{EdS}}\right),\label{T0EdS} \\ 
H_0^\mathrm{EdS}&=&\frac{H_m}{\left(1+z_m^\mathrm{EdS}\right)^{3/2}}, \label{H0EdS}
\end{eqnarray}
\end{subequations}
where $T_0=2.726\pm0.001$ K is the CMB temperature observed today
\cite{Fixsen:2009ug}.

The parameters $T_0^\mathrm{EdS}, H_0^\mathrm{EdS},
\Omega_m=1, \Omega_\mathrm{\Lambda}=\Omega_K=0$ and
the baryon fraction $\Omega_\mathrm{b}$ can then be fed into CAMB,
together with the choice of primordial power spectrum, to find the
spectrum of $\mathit{C}_l$s for the effective EdS model. This will be
identical to the spectrum actually observed at the centre of the void,
at large enough values of $l$.

At small values of $l$ however, the void model will create an
integrated Sachs-Wolfe (ISW) signal which is not captured by the
effective EdS model. However, no rigorous calculation has yet been
made of this expected ISW signal. In addition, the void may also in
principle have a different reionization history than that of the
effective EdS model. These two effects mean that at small $l$ the
$\mathit{C}_l$ values calculated using the EdS model will differ, in
both the TT and TE power spectra, from the actual spectrum due to the
void.

In order to account for this, we choose to apply a cutoff at $l=32$ in
both the WMAP-7 TT and TE power spectra. The value $l=32$ is chosen
also to coincide with the switch-over point for the TT power spectrum
at which the WMAP likelihood routine switches between the low-$l$,
gibbs-sampling likelihood estimation technique and the master code for
high-$l$ (for the TE spectrum the switch-over point between
pixel-based analysis and the master code is close by at $l=24$). This
also allows a more direct interpretation of the likelihood
$\mathcal{L}$ in terms of a $\chi^2$ value. We have checked that
increasing the cutoff point does not materially affect our results.

Given that in Ref.~\cite{Hunt:2007dn} it is found that an EdS model
can fit the WMAP data with a primordial power spectrum similar to the
`bump' model we use here, we expect to find a good fit to the CMB with
$h_0^\mathrm{EdS}\equiv H_0^\mathrm{EdS}/(100
\,\mathrm{kms}^{-1}\mathrm{Mpc}^{-1})\simeq0.44$ and
$T_0^\mathrm{EdS}\simeq T_0$. This should be contrasted with the
values of $h_0^\mathrm{EdS}\simeq0.51$ and
$T_0^\mathrm{EdS}\simeq3.4\,\mathrm{K}$ required with a power-law
primordial spectrum \cite{Zibin:2008vk,Moss:2010jx}.

\subsection{Local Hubble rates}
\label{subsection:H0}

It has been claimed in previous studies
\cite{Zibin:2008vk,Moss:2010jx,Biswas:2010xm} that the \emph{local}
Hubble rate of void models that fit the CMB and SNe Ia data
simultaneously must be very low (as low as $h_0\simeq0.45$ in
Refs.~\cite{Zibin:2008vk,Moss:2010jx}) and that this argues against
void models, since the measured local values (at $z<0.1$) are
significantly higher (see Ref.~\cite{Freedman:2010xv} for a
review). This can provide an important discriminant against void
models. Therefore in performing the MCMC analysis, we also fit the
local Hubble rate for the void model $h_0^\mathrm{LTB}\equiv H/100$ to
the Hubble Key Project (HKP) value $h_0=0.72\pm0.08$
\cite{Freedman:2000cf}. We note that there is some variation in the
value of $h_0$ obtained by different groups, ranging from
$h_0=0.623\pm0.06$ \cite{Tammann:2007ge} to $h_0=0.742\pm0.036$
\cite{Riess:2009pu}, and the HKP value lies in between these two.

The SNe\,Ia compilations, WMAP-7 observations and the HKP value for
$h_0$ form the primary datasets that we use to constrain the void
model. We also discuss the fit to some other cosmological data below.

\subsection{Big bang nucleosynthesis}
\label{subsection:BBN}

Our theoretical understanding of the physics of the epoch of big bang
nucleosynthesis allows us to use observations of the abundances of
various elements to constrain the baryon-to-photon ratio,
$\eta=n_\mathrm{b}/n_\gamma$, at that time. Although $\eta$ is
constant with time in FRW spacetime, this is not the case in LTB
spacetime. However, assuming that $\eta$ is spatially constant in the
LTB model (which need not necessarily be true, see
Refs.~\cite{Regis:2010iq,Clarkson:2010ej}), we can use the effective
EdS model model to calculate $\eta$ for the void model. Since both
models share the same early universe physics by construction, we have
\cite{Moss:2010jx}
\begin{equation}
\label{eta}
\eta_{10}\equiv10^{10}\eta=273.9\left(\frac{T_0}{T_0^\mathrm{EdS}}\right)^3\omega_\mathrm{b}\,,
\end{equation}
where
$\omega_\mathrm{b}\equiv\Omega_\mathrm{b}\left(h_0^\mathrm{EdS}\right)^2$. The
inferred primordial abundance of deuterium, together with that of
helium, provides the constraint $5.1\leq\eta_{10}\leq6.5$ at $95\%$
C.L. \cite{Nakamura:2010zzi}.

\subsection{BAO scale}
\label{subsection:BAO}

The BAO data provided by the SDSS collaboration \cite{Percival:2009xn}
are essentially measurements of a feature in the correlation function
of the observed galaxy distribution which is related to the physical
sound horizon at the CMB scale, evolved down to the redshift at which
the measurement is made. The full theory for how these perturbations
should evolve in LTB spacetimes is unknown and a difficult problem, as
the background curvature enters the Bardeen equation for the
gravitational potential. As this may vary significantly over scales of
$\sim150$ Mpc, it can potentially add a significant and as yet unknown
distortion to BAO scales \cite{February:2009pv}.

Nevertheless, some efforts have been made to compare void models to
BAO data under the assumption that the evolution of perturbations does
\emph{not} depend on scale (as is the case in FRW spacetime)
\cite{Zibin:2008vk,Moss:2010jx,Biswas:2010xm}. Under this assumption
the difference between LTB and FRW spacetimes is simply that as
$H_\parallel \neq H_\perp$ for $r>0$, physical length scales in an LTB
spacetime evolve differently in radial and transverse directions at
late times, whereas in FRW models the evolution is isotropic. It is
not clear that this is necessarily a valid assumption to
make. However, in order to provide a comparison with the previous
studies we follow the same prescription as in Ref.~\cite{Moss:2010jx},
while bearing in mind that a better calculation may lead to
significant changes in the results.

The first step is to construct another effective EdS model, referred
to as the ``BAO model'', which shares the same early universe physics as
the void model under consideration, but unlike in the previous case,
need not match the angular diameter distance to the LSS. We choose in
this case to match the central temperatures, $T_0^\mathrm{BAO}=T_0$,
which then provides us with the relation:
\begin{equation}
\label{H0BAO}
H_0^\mathrm{BAO}=\frac{H_m}{\left(1+z_m\right)^{3/2}},
\end{equation}
as unlike in the previous Section, we will now have
$z_m^\mathrm{BAO}=z_m$ (where, as before, we choose $z_m=100$). We
also match the values of $\Omega_\mathrm{b}$ and $\Omega_m=1$
at this redshift to ensure the same early universe physics.

As the BAO model and the LTB void model have, by definition, the same
physics at coordinates
$\left(r\left(z_m\right),t\left(z_m\right)\right)\equiv\left(r_m,t_m\right)$,
and since at this redshift the LTB spacetime is sufficiently close to
FRW, we may conclude that the physical sound horizon in the void model
at time $t_m$ is essentially isotropic and homogeneous, and equal to
the sound horizon $s_\mathrm{p}\left(z_m\right)$ in the BAO model,
which may be calculated according to \cite{Eisenstein:1997ik}:
\begin{equation}
\label{spz}
s_\mathrm{p}\left(z\right)=\frac{44.5\ln\left[9.83/\left(\Omega_mh_0^2\right)\right]}{\left(1+z\right)\sqrt{1+10\left(\Omega_\mathrm{b}h_0^2\right)^{3/4}}}\,\mathrm{Mpc}.
\end{equation} 
In order to evaluate the BAO scales observed at redshift $z$ by an
observer at the centre of the void, we need to evolve this physical
sound horizon scale down to coordinates
$\left(r\left(z\right),t\left(z\right)\right)$ in the LTB
background. Thus the radial and transverse physical BAO scales at
redshift $z$ will be, respectively,
\begin{eqnarray}
  l_\parallel^\mathrm{BAO}\left(z\right)&=&
  s_\mathrm{p}\left(z_m\right)\frac{A^\prime\left(r\left(z\right),t\left(z\right)\right)}{A^\prime\left(r\left(z\right),t_m\right)}, \label{lBAO:r} \\
  l_\perp^\mathrm{BAO}\left(z\right)&=&
  s_\mathrm{p}\left(z_m\right)\frac{A\left(r\left(z\right),t\left(z\right)\right)}{A\left(r\left(z\right),t_m\right)}, \label{lBAO:t}
\end{eqnarray}
and these can in turn be rewritten in terms of the corresponding
redshift and angular intervals:
\begin{eqnarray}
\Delta z\left(z\right)&=&\left(1+z\right)l_\parallel^\mathrm{BAO}\left(z\right)
H_\parallel\left(r\left(z\right),t\left(z\right)\right), \label{deltaz} \\
\Delta \theta\left(z\right)&=&\frac{l_\perp^\mathrm{BAO}}{A\left(r\left(z\right),t\left(z\right)\right)}. \label{deltatheta}
\end{eqnarray}
It is these values of $\Delta z\left(z\right)$ and
$\Delta\theta\left(z\right)$ that are directly measurable. As
mentioned in Ref.~\cite{Moss:2010jx}, the redshift scale $\Delta
z\left(z\right)$ is expected to be a stronger discriminator of void
models than $\Delta\theta\left(z\right)$. However, while estimates of
the radial BAO scale have been
made\cite{Gaztanaga:2008xz,Gaztanaga:2008de}, the statistical
significance of these claims has been questioned
\cite{Kazin:2010nd,MiraldaEscude:2009uz}. Given this ambiguity, we
choose not to use this data; instead we use constraints on the ratio,
\begin{equation}
\label{Theta}
\Theta\left(z\right)\equiv\frac{s_\mathrm{p}\left(z_\mathrm{rec}\right)}{D_V\left(z\right)},
\end{equation}
at redshifts $z=0.2$ and $z=0.35$ from Ref.~\cite{Percival:2009xn},
where $z_\mathrm{rec}$ is the redshift at recombination and
$D_V\left(z\right)$ is an isotropized distance measure. This is given
in a FRW model by \cite{Eisenstein:2005su}
\begin{equation}
\label{Dv}
D_V^\mathrm{FRW}\left(z\right)\equiv\left[\frac{zd^2_A\left(z\right)}{H\left(z\right)}\right]^{1/3}\;,
\end{equation}
where $d_A\left(z\right)$ and $H\left(z\right)$ represent,
respectively, the angular diameter distance and the Hubble rate in the
FRW model.

As the BAO measurements are quoted in terms of the ratio
$\Theta\left(z\right)$, it is necessary to convert them into a form
that can be related to $\Delta z\left(z\right)$ and
$\Delta\theta\left(z\right)$. It can be shown \cite{Biswas:2010xm}
that
\begin{equation}
\label{Q}
Q\left(z\right)\equiv\left(\Delta\theta^2\Delta z\right)^{1/3}=z^{1/3}\Theta\left(z\right),
\end{equation}
and this is the measure we use to compare the void model to the data.

\section{Method}
\label{section:method}

We perform a likelihood analysis using {\small COSMOMC} \cite{Lewis:2002ah} to
generate Markov-Chain-Monte-Carlo (MCMC) chains to estimate confidence
limits on the parameters in fitting the model to the data. For each
void model, specified by $\Omega_\mathrm{in}$, $r_0$ and $H$, we first
calculate the effective parameters $H_0^\mathrm{EdS}$ and
$T_0^\mathrm{EdS}$ as described in Sec.~\ref{subsection:CMB}.  We
fix the optical depth to the LSS to the WMAP-7 value, although this
choice is immaterial as our cutoff at $l=32$ means that the data do
not constrain reionization in any case.
Therefore the full set of parameters for the void model is: local
matter density $\Omega_\mathrm{in}$, void radius $r_0$, local Hubble
value $H$, baryon fraction $\Omega_\mathrm{b}$, and the power spectrum
parameters $\mathcal{P}_0$, $a$, $b$, $c$ and $k_0$. (The cold dark
matter density $\Omega_\mathrm{c}$ is set equal to
$1 - \Omega_\mathrm{b}$ since we are not considering large-scale
structure formation here so do not need to invoke a small component of
hot dark matter to match the SDSS power spectrum of galaxy clustering
as in Ref.\cite{Hunt:2007dn}.) These are fed as inputs to CAMB
\cite{Lewis:1999bs} to generate the CMB spectrum for the model.

Once the output $\mathit{C}_l$ values have been obtained from CAMB, we
can fit the model to the WMAP-7 data \cite{Komatsu:2010fb}, the
SNe\,Ia data (the Constitution \cite{Hicken:2009dk} and Union2
\cite{Amanullah:2010vv} datasets are fitted separately), the HKP value
for the local Hubble rate \cite{Freedman:2000cf}, the BAO data
\cite{Percival:2009xn} and the BBN constraint \cite{Nakamura:2010zzi}.

\begin{figure*}[htbp]
\center
\includegraphics[scale=1.1]{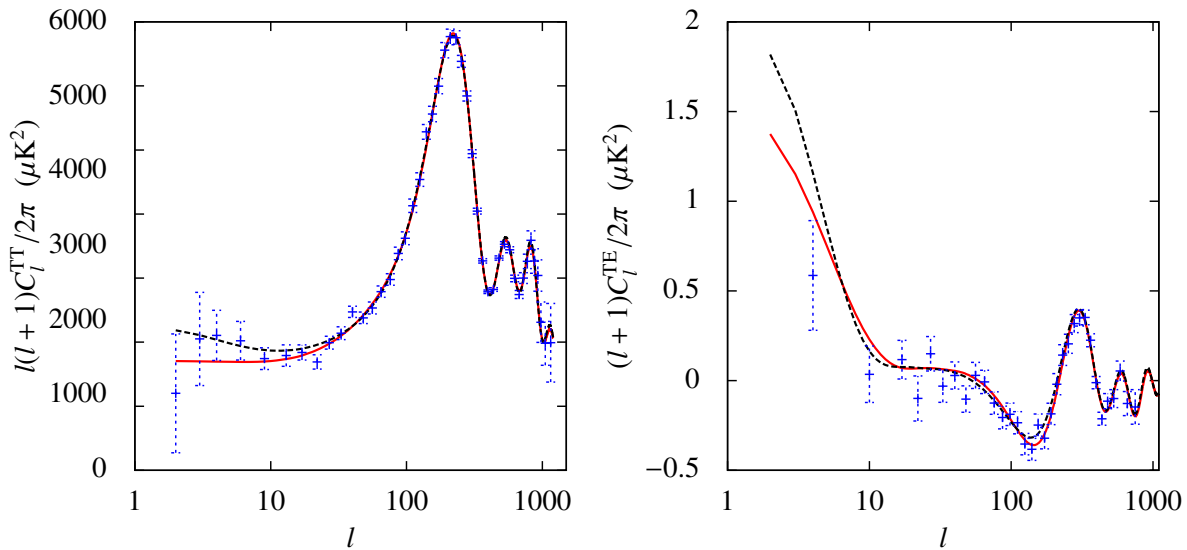}
\caption{\label{figure:Clfig} The TT and TE power spectra of the
  models that best fit the CMB and Constitution SNe\,Ia data. The
  solid (red) curve is for the void model, calculated using the
  effective EdS approach described in Sec.~\ref{subsection:CMB},
  and the dashed (black) curve is for $\Lambda$CDM. Binned WMAP-7 data
  is also shown.}
\end{figure*}

\begin{figure*}[htbp]
\center
\includegraphics[scale=0.42]{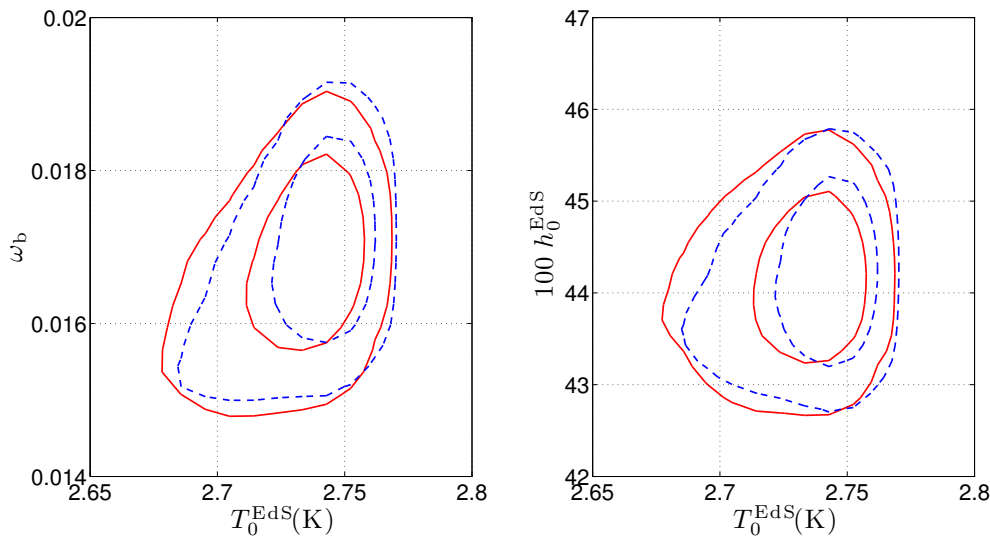}
\caption{\label{figure:2Dmarge} Constraints on the effective EdS
  parameters for the void model with a bump, from CMB and SNe Ia
  data. The solid (red) contours show the $1$ and $2\sigma$ likelihood
  confidence intervals for the Constitution SNe Ia data set and the
  dashed (blue) contours are for Union2.}
\end{figure*}

In order to compare the goodness of fit, we perform exactly the same
fitting procedure for a vanilla $\Lambda$CDM model whose parameters
are: baryon density $\Omega_\mathrm{b}h_0^2$, cold dark matter density
$\Omega_\mathrm{c}h_0^2$, local Hubble rate $100h_0$,\footnote{The
  {\small COSMOMC} code actually uses the ratio of the sound horizon to the
  angular diameter distance as the input parameter, but this is
  equivalent to using $h_0$.} and the power spectrum amplitude, $A_S$,
and slope, $n_s$. We characterize the best-fit likelihood of the void
model by the value
$\Delta\chi^2=-2\ln(\mathcal{L}_\mathrm{void}/\mathcal{L}_{\Lambda\mathrm{CDM}})$,
which means that negative values of $\Delta\chi^2$ favour the void. In
Fig.~\ref{figure:Clfig} the TT and TE power spectra are shown for two
sample best-fit models. The quantitative results of the analysis are
discussed in the next section.

\section{Results}
\label{section:results}

 \begin{figure*}[ht]
\center
\includegraphics[scale=0.45]{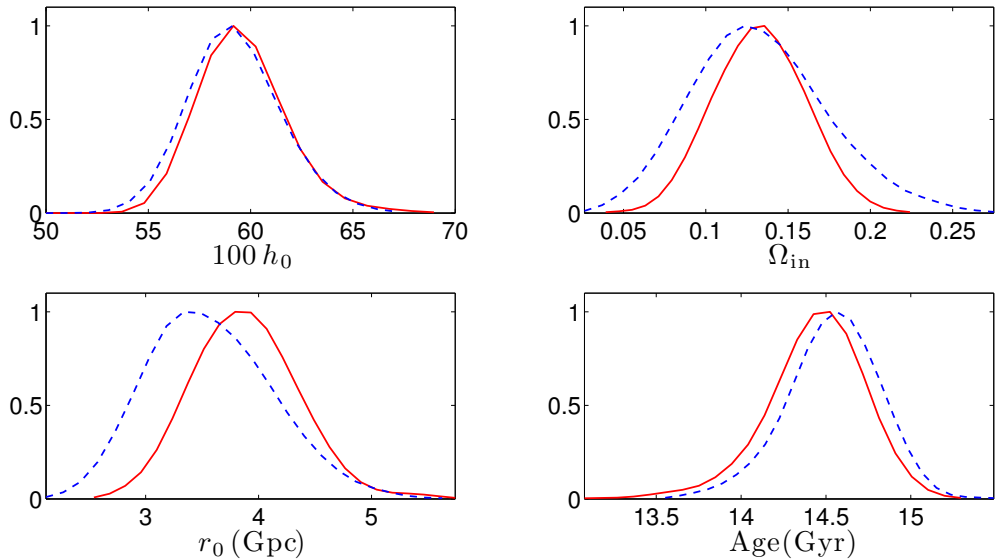}
\caption{\label{figure:1Dmarge} Marginalized 1D likelihoods for the
  void model with a bump, for the fit to CMB and SNe Ia data. The
  solid (red) curves are for the Constitution SNe Ia data set and the
  dashed (blue) curves are for Union2.}
\end{figure*}

In this section we shall use $\mathrm{SN}_\mathrm{CfA}$ to denote the
Constitution data set and $\mathrm{SN}_\mathrm{U2}$ to denote Union2;
CMB denotes the WMAP-7 data and HKP the Hubble Key Project value for
the Hubble rate.

In Figure~\ref{figure:2Dmarge} we plot the 2D likelihoods for the
effective EdS model parameters $T_0^\mathrm{EdS}$, $H_0^\mathrm{EdS}$
and $\omega_\mathrm{b}$ obtained from the MCMC chains for both
CMB+$\mathrm{SN}_\mathrm{CfA}$ and CMB+$\mathrm{SN}_\mathrm{U2}$
data. Clearly, $T_0^\mathrm{EdS}\simeq T_0$ and
$h_0^\mathrm{EdS}\simeq 0.45$, which is exactly as we expect, given
the results of Ref.~\cite{Hunt:2007dn}.  Figure~\ref{figure:1Dmarge}
shows the marginalized likelihoods of the void parameters and the
resultant age of the universe, for the same datasets.

 For both CMB+$\mathrm{SN}_\mathrm{CfA}$ and
 CMB+$\mathrm{SN}_\mathrm{U2}$ we obtain a local Hubble value
 $h_0\sim0.60\pm0.02$, which, although slightly low, is still
 consistent with the HKP value of $h_0=0.72\pm0.08$ to within
 $2\sigma$. It is also much higher than the value of $h_0$ obtained in
 previous studies \cite{Zibin:2008vk,Moss:2010jx}. This is because in
 order to fit the CMB with a power-law primordial spectrum a value of
 $T_0^\mathrm{EdS}\simeq3.4$ K is required and in order to generate
 this, the void must be surrounded by a large over-dense shell. The
 presence of this shell means the asymptotic value $H_0(r\gg r_0)\sim
 H_0^\mathrm{EdS}\sim H_0(0)$ and hence in order to fit the CMB,
 $H_0(0)$ must be small. In contrast, our void profile does not have
 an over-dense shell, and as shown in Figure~\ref{figure:2Dmarge},
 $T_0^\mathrm{EdS}\simeq T_0$. This allows us to generate the
 difference between $h_0\simeq 0.60$ and $h_0^\mathrm{EdS}\simeq 0.45$.

 Our void model also does not suffer from the ``old age'' problem
 referred to in Ref.\cite{Moss:2010jx}: we find the age of the void
 universe to be $14.4\pm0.3$ Gyr for the
 CMB+$\mathrm{SN}_\mathrm{CfA}$ chains and $14.5\pm0.3$ Gyr for
 CMB+$\mathrm{SN}_\mathrm{U2}$, which are significantly less than the
 values of $\sim18.8$ Gyr quoted in \cite{Moss:2010jx}. This is
 unsurprising as the age $t_0$ is related to the value of $h_0$ by
 Eq.~\ref{H0}, and reasonable values for one ensure reasonable
 values for the other.
 
 \begin{table}[htb]

 \caption{\label{table:chi2} Best-fit likelihood values for $\Lambda$CDM and
 the relative $\Delta\chi^2$ values for the void model with a spectral `bump 
for different choices of the fundamental constraining data sets. Here  
$\mathrm{SN}_\mathrm{CfA}$ refers to the Constitution sample and 
$\mathrm{SN}_\mathrm{U2}$ to Union2. CMB refers to WMAP-7 data and HKP to the 
Hubble Key Project value for $h_0$. Constitution data favour the void slightly 
and Union2 data favour $\Lambda$CDM slightly, but neither difference is 
significant given the number of degrees of freedom.}

\begin{ruledtabular}
\begin{tabular}{cccc}
  Datasets&$\#$ d.o.f.&$-2\ln(\mathcal{L}_{\Lambda\mathrm{CDM}})$&$\Delta\chi^2$\\ \hline
 CMB &$1936$&$5785.0$ &$+0.9$ \\
  CMB + $\mathrm{SN}_\mathrm{CfA}$ & 2333 & 6250.4 & $-3.3$\\
  CMB + $\mathrm{SN}_\mathrm{U2}$ & 2493 & 6315.7 & $+1.4$\\
  CMB + $\mathrm{SN}_\mathrm{CfA}$ + HKP & 2334 & 6250.5 & $-0.6$\\
  CMB + $\mathrm{SN}_\mathrm{U2}$ + HKP & 2494 & 6315.9 & $+3.4$\\
\end{tabular}
\end{ruledtabular}
\end{table}

 To enable a comparison of the quality of the fits, we show in
 Table~\ref{table:chi2} the $\Delta\chi^2$ values for the best-fit
 model with a void and a spectral ``bump'', relative to the standard
 $\Lambda$CDM model, for different combinations of the primary
 constraining data sets. It is clear that the $\Delta\chi^2$ values in
 all cases are small given the number of degrees of freedom involved,
 which shows that the void model is perfectly compatible with these
 datasets. A more quantitative comparison cannot be made because as
 noted in Sec.~\ref{subsection:SN} the SNe Ia data are adjusted to
 fit a fiducial $\Lambda$CDM model and the error bars are tuned to
 give a $\chi^2$ per degree of freedom of order unity for this model.

It should be noted that the addition of the HKP constraint increases
the $\Delta\chi^2$ value by $\sim2$ for Union2 and $\sim2.7$ for
Constitution, which reflects the fact that CMB and SNe Ia data favour
a value of $h_0$ that is consistent with, but lower than, the HKP
value. Even with this increase however, the $\Delta\chi^2$ values are
still small (and in fact the void provides a marginally better fit to
the CMB+$\mathrm{SN}_\mathrm{CfA}$ data than does $\Lambda$CDM). It is
possible that a thorough exploration of possible void profiles or use
of unbiased SNe Ia data will make this fit better, but as this does
not detract from the main results presented here, we do not
investigate it in this paper.

A more interesting constraint comes from Big Bang
nucleosynthesis. From our MCMC chains for
CMB+$\mathrm{SN}_\mathrm{CfA}$ and CMB+$\mathrm{SN}_\mathrm{U2}$ we
find in both cases $\eta_{10}=4.6\pm0.2$, which is $\sim2\sigma$ below
the best-fit value. This arises primarily because $\omega_\mathrm{b}$
is on the low side, as seen in Figure~\ref{figure:2Dmarge}. When the
BBN constraint is added to the MCMC chains, we find an increase in
$\Delta\chi^2$ of $\sim2.5$ relative to the values for
CMB+$\mathrm{SN}_\mathrm{CfA}$+HKP and an increase of $\sim3.1$ for
CMB+$\mathrm{SN}_\mathrm{U2}$+HKP. It should be noted however that the
inferred primordial abundance of lithium indicates a significantly
lower value of $\eta$ than does deuterium \cite{Nakamura:2010zzi},
hence a better understanding of the chemical evolution of these
fragile elements is required before the significance of the marginal
discrepancy above can be assessed.

Finally, we find that adding the BAO data results in an increase in
$\Delta\chi^2$ of $4.9$ relative to the value for
CMB+$\mathrm{SN}_\mathrm{U2}$+HKP+BBN and $5.5$ relative to the value
for CMB+$\mathrm{SN}_\mathrm{CfA}$+HKP+BBN, for an additional two
degrees of freedom. The main reason for this increase is that the BAO
data require a shallower void with $\Omega_\mathrm{in}\sim0.17$,
whereas the CMB+SN data favour a slightly deeper void with
$\Omega_\mathrm{in}\sim0.13$. Most of the increase in $\chi^2$ arising
from including the BAO data comes from the poorer fit to SNe Ia data
that results. Given the uncertainty in interpreting
$\chi^2_\mathrm{SN}$ that we have already mentioned, as well as the
assumptions that have gone into our analysis in
Sec.~\ref{subsection:BAO}, we cannot draw any definite conclusions
from these results. However it would appear that the BAO data are
certainly not inconsistent with a void model.

\section{Discussion}
\label{section:discussion}

An interesting proposed test of void models is the Compton
$y$-distortion of the CMB spectrum that is produced by the scattering
of photons by reionized gas in regions of the void that see a highly
anisotropic LSS \cite{Caldwell:2007yu}. In the single-scattering and
linear approximations and under the assumption that the dipole
anisotropy dominates the distortion, this can be written as
\cite{Moss:2010jx}
\begin{equation}
\label{y}
y = \frac{7}{10}\int_0^{z_\mathrm{re}} 
 \mathrm{d}z\frac{\mathrm{d}\tau}{\mathrm{d}z}\beta\left(z\right)^2\,,
\end{equation}
where $\tau$ is the optical depth, $\beta\left(z\right)$ is the dipole
temperature anisotropy in the CMB observed at redshift $z$, and the
integral is taken up to the redshift of reionization $z_\mathrm{re}$.

The FIRAS instrument on COBE provides an upper bound
$y<1.5\times10^{-5}$ (at 2$\sigma$) \cite{Fixsen:1996nj}. While this can constrain some
void profiles, voids lacking an overdense outer shell are found not to be in
tension with this upper bound \cite{Moss:2010jx}. The Gaussian void profile that we consider
does not have a prominent overdense outer shell. Since we do not expect an
interesting constraint, we chose not to evaluate the $y$-distortion,
which would be uncertain in any case since the exact reionization
history (and thus $z_\mathrm{re}$) of the void model is unknown.

Another direct test of a local void is via the kinetic
Sunyaev-Zel'dovich effect \cite{Sunyaev:1972eq}. This has been applied
to cluster kSZ observations (\emph{e.g.},
Ref.\cite{GarciaBellido:2008gd,Yoo:2010ad}) and to full-sky
data \cite{Zhang:2010fa} from the Atacama Cosmology Telescope (ACT) \cite{Das:2010ga}, and is found to place constraints on large voids. Allowing an inhomogeneous bang-time or correctly accounting for the effects of radiation (see \cite{Regis:2010iq,Clarkson:2010ej}) may potentially weaken these constraints while preserving the fits described here. Such an analysis is beyond the scope of the current paper but will be investigated in the future.

In summary, in this paper we have presented a local void model which fits SNe Ia and CMB data, local $H_0$ values, nucleosynthesis constraints and BAO \emph{without} requiring dark energy and thus provides a counterexample to the claim that dark energy is necessary to fit these observations.

\section{Acknowledgements}
\label{section:acknowledgements}

We would like to thank Shaun Hotchkiss, Timothy Clifton, Pedro
Ferreira, Joe Zuntz and Jim Zibin for helpful discussions and correspondence, and Paul Hunt for his early involvement with this paper. SN is supported by an ORS award and
a Clarendon Domus A scholarship from the University of Oxford and
Merton College, Oxford.

\end{document}